# Iterative Reconstruction for Low-Dose CT using Deep Gradient Priors of Generative Model

Zhuonan He#, Yikun Zhang#, Yu Guan, Shanzhou Niu, Yi Zhang, *Senior Member, IEEE*, Yang Chen\*, *Senior Member, IEEE*, Qiegen Liu\*, *Senior Member, IEEE*

*Abstract*——Dose reduction in computed tomography (CT) is essential for decreasing radiation risk in clinical applications. Iterative reconstruction is one of the most promising ways to compensate for the increased noise due to reduction of photon flux. Rather than most existing prior-driven algorithms that benefit from manually designed prior functions or supervised learning schemes, in this work we integrate the data-consistency as a conditional term into the iterative generative model for low-dose CT. At the stage of prior learning, the gradient of data density is directly learned from normal-dose CT images as a prior. Then at the iterative reconstruction stage, the stochastic gradient descent is employed to update the trained prior with annealed and conditional schemes. The distance between the reconstructed image and the manifold is minimized along with data fidelity during reconstruction. Experimental comparisons demonstrated the noise reduction and detail preservation abilities of the proposed method.

## I. Introduction

X-RAY computed tomography (CT) is a popular imaging modality with applications in biology, medicine and other fields. The extensive use of CT examination has raised concerns about the potential risks of carcinogenesis or genetic damage from X-ray radiation [1]. Low-dose CT (LDCT) can image many clinical indications to minimize radiation-related risks without significantly affecting screening or diagnostic performance. Hence, making the radiation dose as low as reasonably achievable is commonly recognized, and it has been a hot research topic during the latest three decades. However, reducing the radiation dose will increase data noise and introduce artifacts into the reconstructed images, which adversely affects its diagnostic performance if these issues are not addressed [2].

Later than the classical filtered back-projection (FBP) algorithm, iterative reconstruction methods become the mainstream in the past few decades [3]-[7]. Incorporating photon statistics and prior information of the target image, these methods have great potentials in noise reduction and information preservation. Specifically, most of the priors are manually designed under a set of neighboring pixels in the image or transform domain, emphasizing on enhancing image smoothness while maintaining edges. Total variation (TV), sparsity in wavelet transform (WT), and non-local patch-based priors have shown promising results in LDCT [5]-[7]. Tirer and Giryes [68] proposed a plug-and-play algorithm that achieves good results by iterating between back-projections and strong denoisers. Nevertheless, these reconstruction approaches may still lose some image details and suffer from remaining artifacts.

Deep learning (DL) techniques, particularly convolutional neural networks (CNN), have been actively developed and applied to various applications recently. The explosive development of them suggests new thinking and huge potential for the medical imaging field. Broadly speaking, these approaches can be categorized into two types: One is using end-to-end supervised DL techniques as a post-processing method [8]-[14], and the other is integrating DL-driven prior techniques into an iterative scheme. In the first class, Chen *et al.* [8] and Kang *et al.* [9] are the first attempt to study the deep CNN for LDCT. By recognizing that the WT operator is able to improve the denoising efficiency and preserves or even enhances the edge features, the results in [10] demonstrated that the directional wavelet utilizing deep CNN was more effective in getting rid of low-dose related noises. Although these algorithms attained promising results, they were usually designed for a particular task with specialized architectures or loss functions, and trained with paired data by taking one image as input and the other as supervision. An efficient fashion to extend the usage of supervised DL is the self-supervised method. It can be used in more practiced applications that the ground truth is not available [66][67].

In another class of methods, it reuses the knowledge in learned priors to tackle various tasks without retraining or modification. As expected, it is possible to model the nonlinear manifold, so that knowledge of normal-dose image can be learned more precisely. Subsequently, the goal of improved reconstruction quality is achieved. Specifically, Baguer *et al.* [15] used deep image priors for CT reconstruction to achieve promising results in the low-data regime. Meanwhile, based on the feature learning and mapping ability of the generative models, many CT reconstruction methods have been proposed from the perspectives of network structure and objective function. Recent progress is mainly driven by two approaches: likelihood-based methods [16]-[19] and generative adversarial network (GAN) [20]. The former uses log-likelihood (or a suitable surrogate) as the training objective, while the latter uses adversarial training to minimize *f*-divergences [21] or integral probability metrics between model and data distribution [22], [23]. For example,

This work was supported in part by the National Natural Science Foundation of China under 61871206, 61661031, 11701097, Science and Technology Program of Jiangxi Province (20192BCB23019, 20202BBE53024) and project of innovative special funds for graduate students in Jiangxi province (CX2019075). (Z. He and Y. Zhang are co-first authors.) (Corresponding authors: Yang Chen; Qiegen Liu.)

Z. He, Y Guan and Q. Liu are with the Department of Electronic Information Engineering, Nanchang University, Nanchang 330031, China. ({hezhuonan, guanyu}@email.ncu.edu.cn, {liuqiegen}@ncu.edu.cn).

S. Niu is with the School of Mathematics and Computer Science, Gannan Normal University, Ganzhou 341000, China. (szniu@gnnu.edu.cn).

Y. Zhang is with the College of Computer Science, Sichuan University, Chengdu, Sichuan 610065, China. (yzhang@scu.edu.cn).

Y. Chen, and Y. Zhang are with the Laboratory of Image Science and Technology, School of Computer Science and Engineering, Southeast University, Nanjing 210096, China. ({yikun, chenyang.list}@seu.edu.cn).

Adler *et al.* [24] employed a Wasserstein GAN to draw samples from the conditional distribution. Regretfully, despite the success of generative models in tackling natural images, there have been few studies in field of medical imaging, especially in LDCT.

In this work, to boost the effectiveness of LDCT reconstruction, we focus on exploring dEep grAdient priorS of genErative modeL (EASEL). As a newly developed unsupervised learning, the score-based generative model has exhibited great potential for diverse image representation [25]. Viewing the noisy observation as a conditional variable, a conditional score-based generative model is introduced into LDCT reconstruction in this work. Since EASEL is implemented in a fully unsupervised way, it has more flexibility and less requirement on training data [26].

The major contributions of this paper are:
- *Estimating the gradient of data density as a prior:* Unlike the traditional generative models that estimate the data density as priors directly, we utilize the gradient of data density as a prior. This strategy will be more suitable for iterative reconstruction procedure. As well known, the basic operator in iterative algorithm is to calculate the gradient of the objective function. When the gradient of data density is learned, it can be easily incorporated into iterative algorithms.
- *Flexibility of the iterative reconstruction scheme:* In the framework of the iterative scheme, two strategies are employed: Conditional and annealed strategies. First, based on Bayesian rule, the data consistency is incorporated into the loop of prior generation procedure. Second, via annealing strategy, the stochastic gradient descent (i.e., Langevin dynamics) is employed to enable the gradient of data density gradually approach to the "true" prior.

The remainder of the paper is organized as follows. Section II provides a brief description of preliminary work with regard to the generative model and gradient of generative model. Section III presents forward formulation of LDCT and the corresponding iterative solver. Extensive experimental comparisons between the proposed EASEL and state-of-the-arts are conducted in Section IV. Finally, concluding remarks and directions for future research are given in Section V.

## II. PRELIMINARIES

### A. Deep Generative Models

Recently advances with deep generative networks have shown promising results in modeling complex distributions such as images [27], audios [28] and texts [29]. As shown in Fig.1, the popular deep generative models can be primarily categorized into two groups: Explicit generative model and implicit generative model. The former model provides an explicit parametric specification of the data distribution, specifying a log-likelihood function $\log p(x)$, including autoencoders (AE) and its variants [30], [31], flow-based generative models [32], [33], and deep Boltzmann machine [34]. Specially, score-based models [37]-[43] train parametric network to approximate the likelihood gradient $\nabla_x \log p(x)$, which have tractable likelihood estimation. Alternatively, we can specify implicit probabilistic models that define a stochastic procedure to directly generate data. GANs [16] are the well-known implicit likelihood models, in the sense that they optimize the objection function through adversarial learning, and have been shown to produce high quality images [20], [35], [36]. Despite its success, GANs still suffer from a remarkable difficulty in training and interpretability due to the lack of theory guarantee.

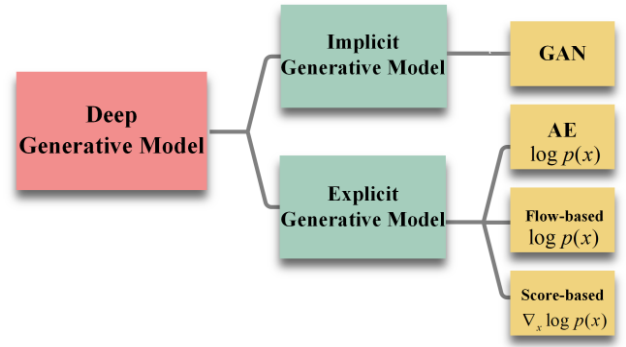

**Fig. 1.** Classification of the popular deep generative models.

### B. Deep Gradient Priors of Generative Model

Recent works show that successful image generation can be achieved by score-based generative models [37]-[43]. They represent probability distributions through score functions $\nabla_x \log p(x)$ —a vector field pointing in the direction where the likelihood of data $p(x)$ increases the most rapidly. Remarkably, these score functions can be learned from data without requiring adversarial optimization, and can produce realistic image samples that rival GAN [37]. To facilitate the description of the present method, we tabulate the notations used hereafter in Table I.

TABLE I SUMMARIZATION OF NOTATIONS.

| Notation | Description | Notation | Description |
|---|---|---|---|
| $x$ | Data sample | $x_0$ | Initial sample |
| $p(x)$ | Likelihood of data | $A$ | System matrix |
| $\nabla_x \log p(x)$ | Gradient of the data density | $y$ | Observation data |
| $s_\theta(x)$ | Score network with parameter $\theta$ | $b_i$ | X-ray source intensity |
| $\{\sigma_l\}_{l=1}^{L}$ | Noise with different scale magnitudes | $r_i$ | Imaging noise |
| $\varepsilon$ | Step size for sample update | $N_m$ | The number of measurements |
| $T$ | Number of iterations under a single noise scale | $N_v$ | The number of image voxels |
| $p_{\sigma_l}(x)$ | The distribution of $x+\eta$ | $p(x|y)$ | Posterior data distribution |
| $L$ | Number of noise scales | $\beta=\beta(\lambda)$ | Penalty parameter |
| $I_i$ | Number of transmitted photons | $\gamma$ | Relaxation factor |

The whole procedure consists of two steps: First, the denoising score matching (DSM) is used to train a network $s_\theta(x)$ to approximate $\nabla_x \log p_{\sigma_l}(x)$ with different scale magnitude $\{\sigma_l\}_{l=1}^L$; Second, via annealing strategy of deceasing $\sigma_l$, the Langevin dynamics is executed to draw sampling from a sequence of $s_\theta(x;\sigma_l)$ and approaches to the final estimation of $\log p(x)$.

***Denoising Score Matching:*** In general, using score matching [41], a score network $s_\theta(x)$ can be trained to estimate $\nabla_x \log(p(x))$ by minimizing the objective function:

$$E_{p(x)}[\|s_\theta(x) - \nabla_x \log p(x)\|_2^2]. \tag{1}$$

Subsequently, Vincent [42] used DSM $s_\theta(x)$ to match a non-parametric kernel density estimator:

$$E_{p_\sigma(\tilde{x})}[\|s_\theta(\tilde{x}) - \nabla_{\tilde{x}} \log p_\sigma(\tilde{x})\|_2^2], \tag{2}$$

where the corresponding perturbed data distribution is $p_\sigma(\tilde{x}) = \int p_\sigma(\tilde{x}|x)p(x)dx$. Crucially, one caveat of DSM is that the optimal score $s_{\theta^*}(x) = \nabla_x \log p_\sigma(x) \approx \nabla_x \log p(x)$ is true only when the noise $\sigma$ is small enough. Whereas, learning the score function with the single-noise perturbed data distribution will lead to inaccurate score estimation in the low data density region on high-dimension data space, which could be severe due to the low-dimensional manifold assumption. Thus, Song and Ermon [25] proposed learning a single neural network based on multiple perturbed data distributions with Gaussian noises of varying magnitudes:

$$\frac{1}{2L}\sum_{l=1}^L \sigma_l^2 E_{p_{\sigma_l}(\tilde{x}|x)p(x)}[\|s_\theta(\tilde{x};\sigma_l) - \nabla_x \log p_{\sigma_l}(\tilde{x}|x)\|_2^2], \tag{3}$$

where the target $\nabla_x \log p_{\sigma_l}(\tilde{x}|x) = (x-\tilde{x})/\sigma_l^2$ has a simple closed-form and the empirical average is utilized to estimate all expectations.

***Annealed Langevin Dynamics:*** In many situations, score function is easier to model and estimate than the original density function [43]. For example, for an unnormalized density it does not depend on the partition function. Once the score function is known, we can employ Langevin dynamics to sample from the corresponding distribution. The Langevin dynamic algorithm is an efficient Markov chain Monte Carlo sampling method. Given a step size $\varepsilon > 0$, a total number of iterations $T$, an initial sample $x_0$, it evaluates the gradient of the negative log-probability iteratively:

$$\begin{aligned} x^t &= x^{t-1} + \frac{\varepsilon}{2}\nabla_x \log p_{\sigma_l}(x^{t-1}) + \sqrt{\varepsilon}z^{t-1} \\ &= x^{t-1} + \frac{\varepsilon}{2}s_\theta(x^{t-1};\sigma_l) + \sqrt{\varepsilon}z^{t-1} \end{aligned}, \tag{4}$$

where $z \sim N(0,I)$ and $t = 1,\cdots,T$.

However, when two modes of the data distribution are separated by low density regions, Langevin dynamics will not be able to correctly recover the relative weights of these two modes in reasonable time, and therefore the sampling model might not converge to the true distribution. In addition, since Langevin dynamics is usually initialized in low-density regions of the data distribution, inaccurate score estimation in these regions will negatively affect the sampling process. Hence, mixing can be difficult because of the need of traversing low-density regions to transition between modes of the distribution. To tackle these two challenges, Song and Ermon [25] proposed an annealed version of Langevin dynamics, where they initially used scores corresponding to the highest noise level, and gradually annealed down the noise level $\{\sigma_l \to 0\}_{l=1}^L$ until it was small enough to be indistinguishable from the original data distribution. As a remedy, large noise levels will produce samples in low density regions of the original data distribution, which can improve score estimation.

## III. METHODOLOGY

### A. LDCT Imaging Model

The statistics of CT measurement data are often described by a Poisson distribution [44]. Specifically, a Poisson model for the intensity measurement is

$$I_i \sim Poisson\{b_i e^{-[Ax]_i} + r_i\}, \quad i = 1,\cdots,N_m, \tag{5}$$

where $I_i$ is the number of transmitted photons, $A$ is a system matrix (projection operator), $b_i$ denotes the X-ray source intensity of the $i$-th ray, and $r_i$ denotes the background contributions of scatter and electrical noise. $x$ is a vector for the representation of attenuation coefficients with units of inverse length, $N_m$ is the number of measurements and $N_v$ is the number of image voxels.

After taking the logarithm operation, the sinogram data are often approximated as a weighted Gaussian [10]:

$$y_i \sim N([Ax]_i, \bar{I}_i/(\bar{I}_i - r_i)^2), \tag{6}$$

where $\bar{I}_i = E[I_i]$. In fact, the LDCT problem can be formulated as an inverse problem $y = Ax + \mu$. In order to cover the uncertainties that occurs especially with ill-posed problems, the theory of Bayesian inversion considers the posterior distribution $p(x|y)$ [45]. This posterior is the conditional density of the image $x$ conditioned on the measurements $y$.

### B. Proposed EASEL Model

In this paper, we are devoted to using the generative model to tackle the LDCT problem, a process of noise addition in imaging [52]. Accordingly, the estimation problem can be reduced to

$$\begin{aligned} \hat{x} &= \arg\max_x \log(p(x)) \\ s.t. \quad y &= Ax + \mu \end{aligned}. \tag{7}$$

Owing to the observation $y$, the sampling object is not directly from $p(x)$, but the posterior distribution $p(x|y)$, i.e.,

$$x^t = x^{t-1} + \frac{\varepsilon}{2}\nabla_x(\log p(x^{t-1}|y)) + \sqrt{\varepsilon}z^{t-1}. \tag{8}$$

Due to the Bayesian rule $p(x|y) = p(y|x)p(x)/p(y)$, it becomes to be

$$\begin{aligned} x^t &= x^{t-1} + \frac{\varepsilon}{2}\nabla_x[\log p(x^{t-1}) + \log p(y|x^{t-1})] + \sqrt{\varepsilon}z^{t-1} \\ &= x^{t-1} + \frac{\varepsilon}{2}\nabla_x(\log p(x^{t-1})) + \frac{\varepsilon\lambda}{2}\nabla_x[\|y - Ax^{t-1}\|^2] + \sqrt{\varepsilon}z^{t-1} \end{aligned}. \tag{9}$$

Here, $\log p(y|x)$ is given by the data model that is usually derived from knowledge about how data is generated and $\log(p(x))$ is given by the prior model that represents information known beforehand about the true model parameter. The hyperparameter $\lambda$ balances the trade-off between priors and data fidelity. It has to be estimated if we do not know the standard deviation of the measurement noise.

Given the gradient of data density $\nabla_x(\log p(x))$ to be learned in advance, Eq. (9) is essentially a stochastic gradient descent scheme with conditional strategy. Besides of the conditional strategy, in this study we simultaneously use an annealed strategy along the iterative procedure, i.e., $\nabla_x(\log p_\sigma(x)) \to \nabla_x(\log p(x))$ with artificially decreasing $\sigma$-value (in Subsection III.C). A geometric interpretation of the whole iterative procedure is visualized in Fig. 2. Overall, the conditional strategy guarantees that the intermediate solution will approach to the data-feasible domain. At the meantime, the annealing strategy tries its best to pursue the optimal prior knowledge in a solid manner [73]. It is predicable that global-scale structures discovered at high noise level will be preserved as $\sigma$-value drops, meanwhile the occurrence of fine-detail structures is still admitted.

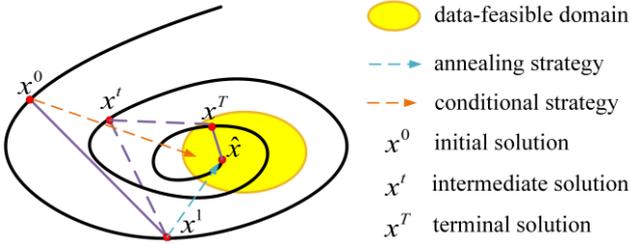

**Fig. 2.** A geometric interpretation of the present EASEL algorithm. The optimization would minimize the distance between the feasible domain and the manifold.

### C. Optimization for EASEL

In the following, we elaborate on the details for implementing Eq. (9).

First, we use annealing strategy to approximate Eq. (9). Let $p_{\sigma_l}(x)$ denote the distribution of $x+\eta_{\sigma_l}$ for $x \sim p$ and $\eta_{\sigma_l} \sim N(0,\sigma_l^2 I)$. At high noise levels $\sigma_l$, $p_{\sigma_l}(x)$ are approximately Gaussian and irreducible, so the Langevin dynamics Eq. (9) will mix quickly. The modified Langevin dynamics is as follows:

$$x^t = x^{t-1} + \frac{\varepsilon}{2}\nabla_x(\log p_\sigma(x^{t-1})) + \frac{\varepsilon\lambda}{2}\nabla_x[\|y-Ax^{t-1}\|^2] + \sqrt{\varepsilon}z^{t-1}. \quad (10)$$

Second, as done in [24][53], we seek to an alternating technology to handle Eq. (10). Especially, we decompose Eq. (10) into two sub-iterative steps as:

$$u^t = x^{t-1} + \frac{\varepsilon_l}{2} s_\theta(x^{t-1};\sigma_l) + \sqrt{\varepsilon_l}z^{t-1}, \quad (11)$$

$$x^t = \arg\min_x[\|y-Ax^{t-1}\|^2 + \beta\|x^{t-1} - u^t\|^2], \quad (12)$$

where parameter $\beta = \beta(\lambda)$ is related to $\lambda$. It is worth noting that the first term $\|y-Ax\|^2$ in Eq. (12) is difficult to be minimized directly. Similarly as done in our previous work [74], the separable quadratic surrogates (SQS) algorithm [54][55] is adopted to get the reconstructed CT volume from the convex problem (12):

$$x_j^k = x_j^{k-1} - \frac{\sum_{i=1}^{M}(A^T(Ax_i^{k-1}-y_i)) + \beta(x_j^{t-1}-u_j^k)}{A^T A\mathbf{1} + \beta\mathbf{1}}, \quad (13)$$

where $A^T$ is the transpose of $A$ and refers to back-projection, 1 is an all-ones vector. Especially, the division is component-wise operator, it makes the algorithm very fast and monotonous in the iterative process.

We begin by initializing $z^0$ as the standard Gaussian vector, and then take a gradient step in one of these while fixing the other. Additionally, for additional acceleration, we apply the Nesterov's momentum [56] that exploits the previous descent directions. A momentum term can be $w^{t+1} = x^{t+1} + \gamma(x^{t+1}-x^t)$, where $\gamma$ is a relaxation factor that lies between 0 and 1.

In summary, the visual flowchart of training phase and iterative reconstruction phase for LDCT is shown in Fig. 3. Furthermore, **Algorithm 1** explains the reconstruction algorithm in detail. EASEL consists of two loops. The outer loop is composed of several stages to enable that $p_{\sigma_l}(x)$ tends to $p(x)$ with decreasing $\sigma_l$-value. The inner loop conducts the conditional Langevin dynamics with alternatingly updating of data-consistency and deep gradient priors. Since both the Langevin dynamics updating [57] and SQS updating [54][55] have convergence guarantee, the overall EASEL algorithm will come to the convergence region after finite iterations.

---

**Algorithm 1** EASEL for Low Dose CT Imaging
---
**Initialization**: $x^0$ and $w^0$, $\sigma \in \{\sigma_l\}_{l=1}^L$, $\beta$, $\gamma$, $\tau$
**For** $l = 1, 2, \cdots, L$
    $\varepsilon_l = \tau\sigma_l^2/\sigma_L^2$
    **For** $t = 1, 2, \cdots, T$
        Draw $z^{t-1} \sim N(0,1)$
        $u^t = x^{t-1} + \frac{\varepsilon_l}{2}s_\theta(x^{t-1};\sigma_l) + \sqrt{\varepsilon_l}z^{t-1}$
        $x^t = w^{t-1} - \frac{A^T(Ax^{t-1}-y) + \beta(x^{t-1}-u^t)}{A^T A\mathbf{1} + \beta\mathbf{1}}$
        $w^t = x^t + \gamma(x^t - x^{t-1})$
    **End for**
    $x^0 \leftarrow w^T$
**End for**

---

### D. Network Architecture of $s_\theta(x,\sigma)$

As described above, the adaption of unsupervised network to the general LDCT reconstruction is the key contribution of EASEL. Besides, the architecture of the score-based network $s_\theta(x,\sigma)$ is also an important factor that impacts the algorithm performance. In this subsection, following the idea of high-dimensional embedded denoising network in our previous work [51][52], we present a channel-copy guided RefineNet as $s_\theta(x,\sigma)$.

Originally, the RefineNet [46] is a modern variant of U-Net [47] that also incorporates ResNet designs [48]. On this basis, Song and Ermon [25] replaced the max pooling layers in refine blocks with multi-path block, as multi-path block is reported to produce smoother and more diversity features for image generation tasks such as image reconstruction. More importantly, they used convolutional operator to produce high-dimensional features before the multi-path block, as seen in Fig. 4(a), such as to obtain flexible representation and excellent robustness abilities.

Alternatively, in Fig. 4(b), we choose the channel-copy strategy to replace the convolutional operator in naïve RefineNet to form the channel-copy guided RefineNet. As seen in Fig. 4, both the naïve RefineNet and channel-copy guided RefineNet extend the generative ability via extending the information/representation dimension, such as increasing channel dimension of the input. However, the latter strategy is simpler and favors to computation effectiveness. As demonstrated in Section IV. E, the naïve RefineNet needs 64 convolutional kernels, while the present channel-copy guided RefineNet only uses 10 coped channels.

Fig. 5 depicts the whole architecture of the channel-copy guided RefineNet. At the prior training stage, we copy and rearrange the single-channel image to the same 10-channel images. Thus, the DSM network can be trained with these data injected with artificial Gaussian noise. At the iterative reconstruction stage, in order to pave the way to apply the trained 10-channel prior to the intermediate single-channel image from the previous iteration, it needs to conduct the channel-copy and channel-mean operators before and after the Langevin dynamics updating.

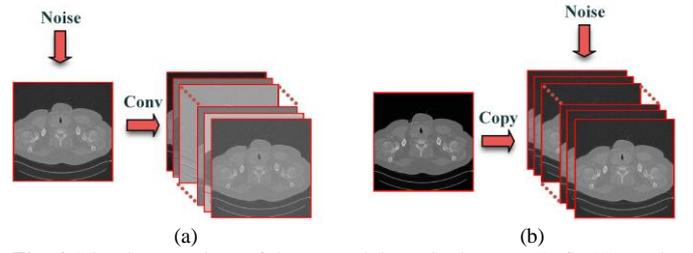

**Fig. 4.** Visual comparison of the network input in the naïve RefineNet and the modified network $s_\theta(x;\sigma)$ used in EASEL.

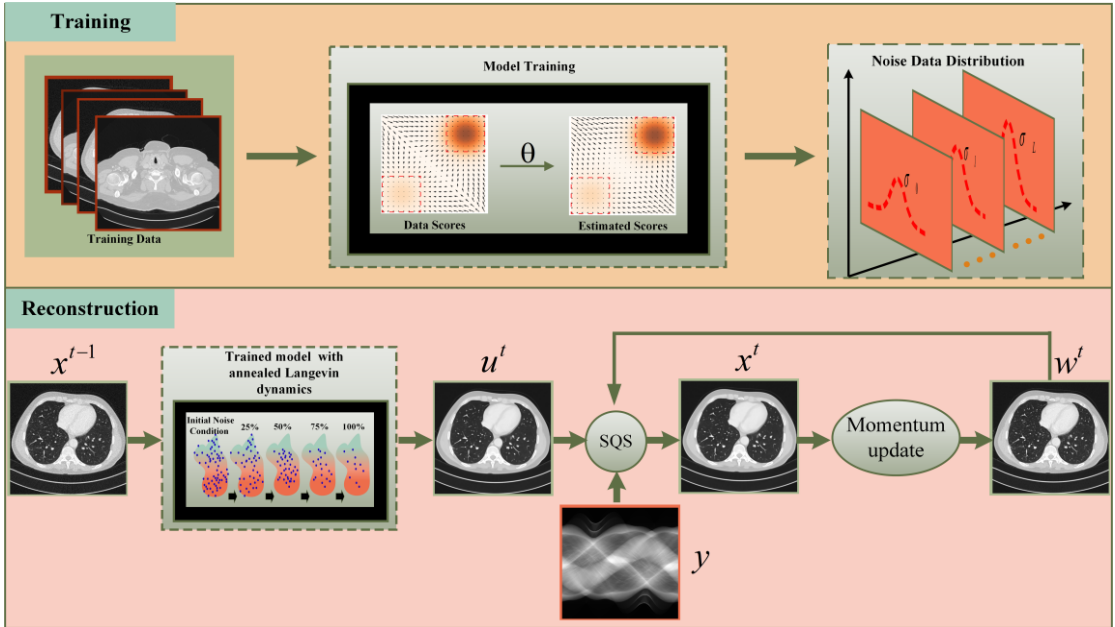

**Fig. 3.** The training and reconstruction paradigm of the generative model-based algorithm EASEL. It consists of two components, i.e., a denoising score matching for score estimation involving various noise magnitudes simultaneously, and an iterative cycle for reconstruction including the annealed and conditional Langevin dynamics.

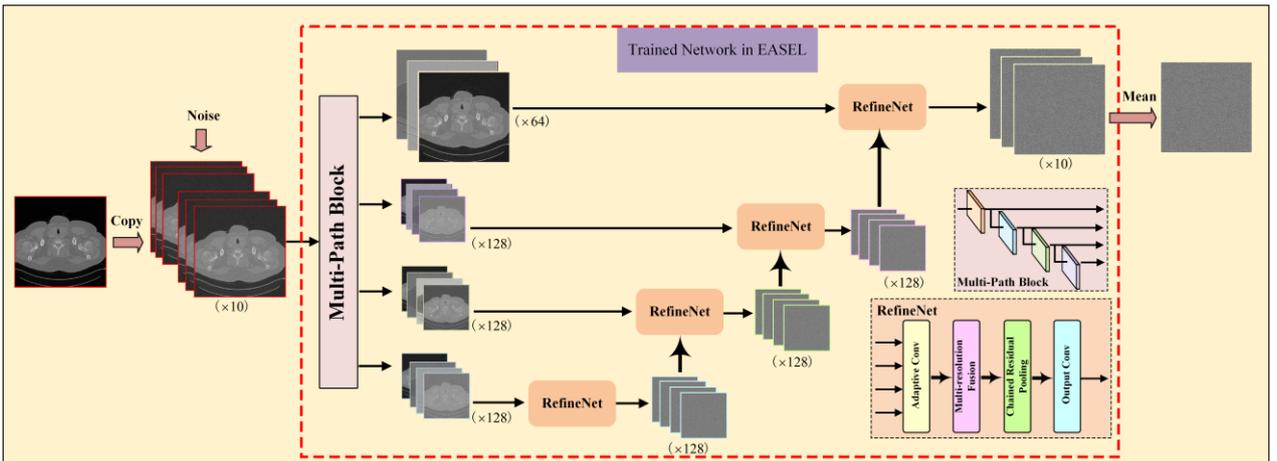

**Fig. 5.** The architecture of channel-copy guided RefineNet used in $s_\theta(x;\sigma)$ of EASEL. A distinct difference of the channel-copy guided RefineNet from the naïve RefineNet is that, we use channel-copy technique to attain high-dimensional features and then inject noise into them. A more detailed visual comparison is shown in Fig. 4.

## IV. EXPERIMENTAL RESULTS

In the experiments, five methods are compared with EASEL including the classical FBP reconstruction (Ramp-filter) [58], TV-based iterative method [59], dictionary learned by K-SVD algorithm [60], RED-CNN network [61] and domain progressive residual network DP-ResNet [62]. The involved parameters are set following the guidelines in their original papers. For more in-depth study and research, the source code of EASEL can be found at: https://github.com/yqx7150/EASEL.

### A. Data Specification

***AAPM Challenge Data:*** To evaluate the algorithm on a clinically realistic use-case, we consider reconstruction of simulated data from human abdomen CT scans as provided by Mayo Clinic for the AAPM Low Dose CT Grand Challenge [63]. The data includes high-dose CT scans from 10 patients, of which we use 9 for training and 1 for evaluation. We use the 1 $mm$ slice thickness reconstructions, resulting in 2961 training images with each 512×512 pixel in size. We use a two-dimensional fan-beam geometry with 1000 angles, 1000 pixels, source to axis distance 500 mm and axis to detector distance 500 mm. The corresponding LDCT images are simulated by adding Poisson noise (whose intensities are $b_i = 5e4$ in Eq. (5)) into the sinogram data [64].

**CIRS Phantom Data:** A high-quality set of CT volumes (512×512×100 voxels, voxel size 0.78×0.78×0.625 mm$^3$) of an anthropomorphic CIRS phantom is obtained from a GE Discovery HD750 CT system, in which the tube current value is set to 600 mAs to guarantee a good image quality for low-dose simulation under 150 mAs. The source-to-axial distance is 573 mm, and the source-to-detector distance is 1010 mm. Fig. 6 displays some representative high-dose images, low-dose CT images and the differential images.

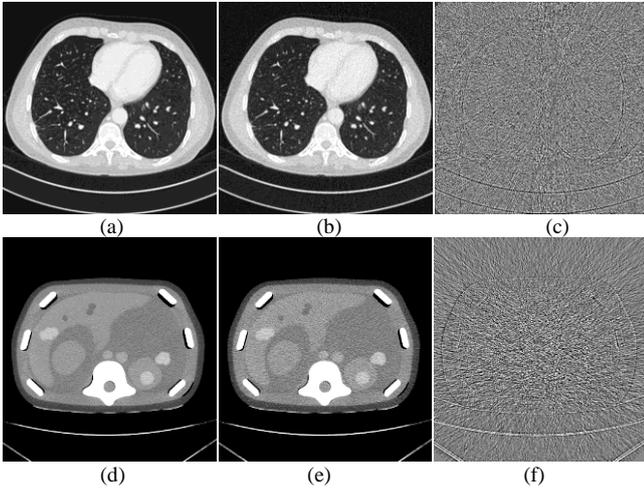

**Fig. 6.** Visual illustration of LDCT data and its counterpart high-dose CT image. (a)(d) High-dose CT in AAPM challenge data and CIRS phantom data, respectively; (b)(e) The corresponding LDCT images; (c)(f) Difference image between high-dose CT and LDCT image.

### B. Model Training and Parameter Selection

In our experiments, the model was trained by the Adam algorithm with the learning rate $10^{-3}$ and Kaiming initialization is used to initialize the weights. The method implemented in Python using Operator Discretization Library (ODL) [72] and PyTorch on a personal workstation with a GPU card (GeForce RTX 2080) accelerated the calculation process. An example of model fitting when training the model is given in Fig. 7, it can be seen that as the number of iterations increases, the model gradually converges and remains stable. In the reconstruction stage, we evaluated several parameter combinations and finalized the parameter settings in as follows. The iteration number is set to $T=150$ for each noise scale and the scale number $L=12$, the relaxation factor $\gamma=0.5$ and regularization weight $\lambda=150$, $\tau=1.8\times10^{-5}$.

For a fair comparison, the parameters of KSVD and TV are hand-tuned for the best performance. As for KSVD method, the size of overlapping patches is set to $8\times8$, the atom number is set to 64, and the tolerance parameter is set to 0.8. Both RED-CNN and DP-ResNet learn an end-to-end mapping from LDCT images to normal-dose images. The LDCT images were simulated by adding Poisson noise with intensities $b_i = 5e4$.

### C. Quantitative Indices

To evaluate the quality of the reconstructed images, three metrics, mean absolute error (MAE), peak signal-to-noise ratio (PSNR) and structural similarity index (SSIM), are selected for quantitative assessment.

Specifically, in statistics, MAE is a measure of errors between paired observations expressing the same phenomenon. It is defined as:

$$MAE = \sum_{i=1}^{N} |x_i - \hat{x}_i| / N, \quad (14)$$

where $N$ is the number of pixels in the reconstructed image. MAE approaches to zero if the reconstructed image is closer to the reference image.

The PSNR measure describes the relationship of the maximum possible power of a signal with the power of noise corruption. Higher PSNR means better image quality. Denoting $x$ and $\hat{x}$ to be the reconstructed image and ground truth, PSNR is expressed as:

$$PSNR(x,\hat{x}) = 20\log_{10}[\text{Max}(\hat{x})/\|x-\hat{x}\|_2]. \quad (15)$$

The SSIM-value is used to measure the similarity between the original CT image and reconstructed images, evaluated on three aspects: luminance, contrast, and structural correlation. SSIM values are normalized between 0 and 1, being 1 the situation in which both images are equal. SSIM is defined as:

$$SSIM(x,\hat{x}) = \frac{(2\mu_x\mu_{\hat{x}} + c_1)(2\sigma_{x\hat{x}} + c_2)}{(\mu_x^2 + \mu_{\hat{x}}^2 + c_1)(\sigma_x^2 + \sigma_{\hat{x}}^2 + c_2)}, \quad (16)$$

where $\mu_x$ and $\sigma_x^2$ are the average and variances of $x$. $\sigma_{x\hat{x}}$ is the covariance of $x$ and $\hat{x}$. $c_1$ and $c_2$ are used to maintain a stable constant.

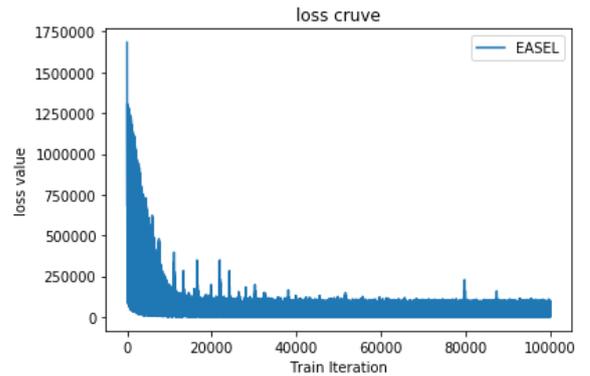

**Fig. 7.** Plot of the loss function across training iterations. The convergence behavior is fast and stable.

### D. Reconstruction Results

**AAPM Challenge Data:** For better evaluation of reconstruction image quality, the MAE, PSNR and SSIM values are given in terms of Means ± STDs (average scores ± standard deviations) for the LDCT images in the test dataset. Table II presents all the results, and the best value of each metric is marked in black bold. Intuitively, our method scores the highest PSNR, and ranks the second in terms of MAE and SSIM values, whereas the DP-ResNet ranks second in PSNR measure. At the same time, comparing the experimental results between TV and K-SVD, both of them obtain worse image quality evaluation metrics. It implies that the performance of the classical algorithms may still lose some details and suffer from remaining artifacts.

To visually illustrate the reconstruction performance, we perform qualitative comparisons over the selected methods for CT images of different body parts, as shown in Fig. 8. It is noteworthy that the results focus on content restoration, artifact suppression, and noise reduction. In addition, for better evaluation of image quality, Fig. 8 depicts the zoomed regions-of-interest (ROI) marked by the red rectangles. From the FBP reconstructions in Fig. 8(b1)-(b3), we can see that FBP reconstruction leads to severely degraded LDCT images with obviously amplified noise and artifacts. As a traditional method, K-SVD produces reconstruction images (Fig. 8(c1)-(c3)) with visually smoother appearances. However, some tiny structures might be smoothed out and lead to lowered tissue contrast, as indicated by the red arrows. Furthermore, it also can be seen that deep learning methods effectively reduce noise and remarkably overmatch TV and K-SVD in Fig. 8(e1)-(f3). They improve the effect of noise reduction and suppress most artifacts. However, they have incomplete preservation of details and texture information.

Comparing the results in Fig. 8(g1)-(g3), we can see that the proposed EASEL method achieves the best performance in terms of noise-artifact suppression and tissue feature preservation. The corresponding residual images are shown in Fig. 9. Among them, we find that TV is able to perform well but tends to smooth textures and edges. Compared to the competitive reconstruction methods, EASEL has its own advantages, which can effectively reduce the noise, and its reconstruction performs better in terms of artifact reduction and detail preservation.

TABLE II
QUANTITATIVE RESULTS (MEAN±STD) ASSOCIATED WITH DIFFERENT METHODS FOR THE IMAGES IN THE TESTING LD PROJECTION DATASET.

| Method | MAE | PSNR | SSIM |
| --- | --- | --- | --- |
| FBP(Ramp-filter) | 67.69±13.16 | 28.68±2.03 | 0.4443±0.0718 |
| TV | 29.65±4.60 | 34.98±1.32 | 0.8610±0.0348 |
| K-SVD | 21.00±1.04 | 35.68±2.26 | 0.8198±0.0241 |
| RED-CNN | 16.97±1.02 | 39.37±1.87 | 0.9478±0.0056 |
| DP-ResNet | **14.97**±0.82 | 41.03±1.46 | **0.9572**±0.0050 |
| EASEL | 15.44±1.17 | **41.58**±0.67 | 0.9539±0.0067 |

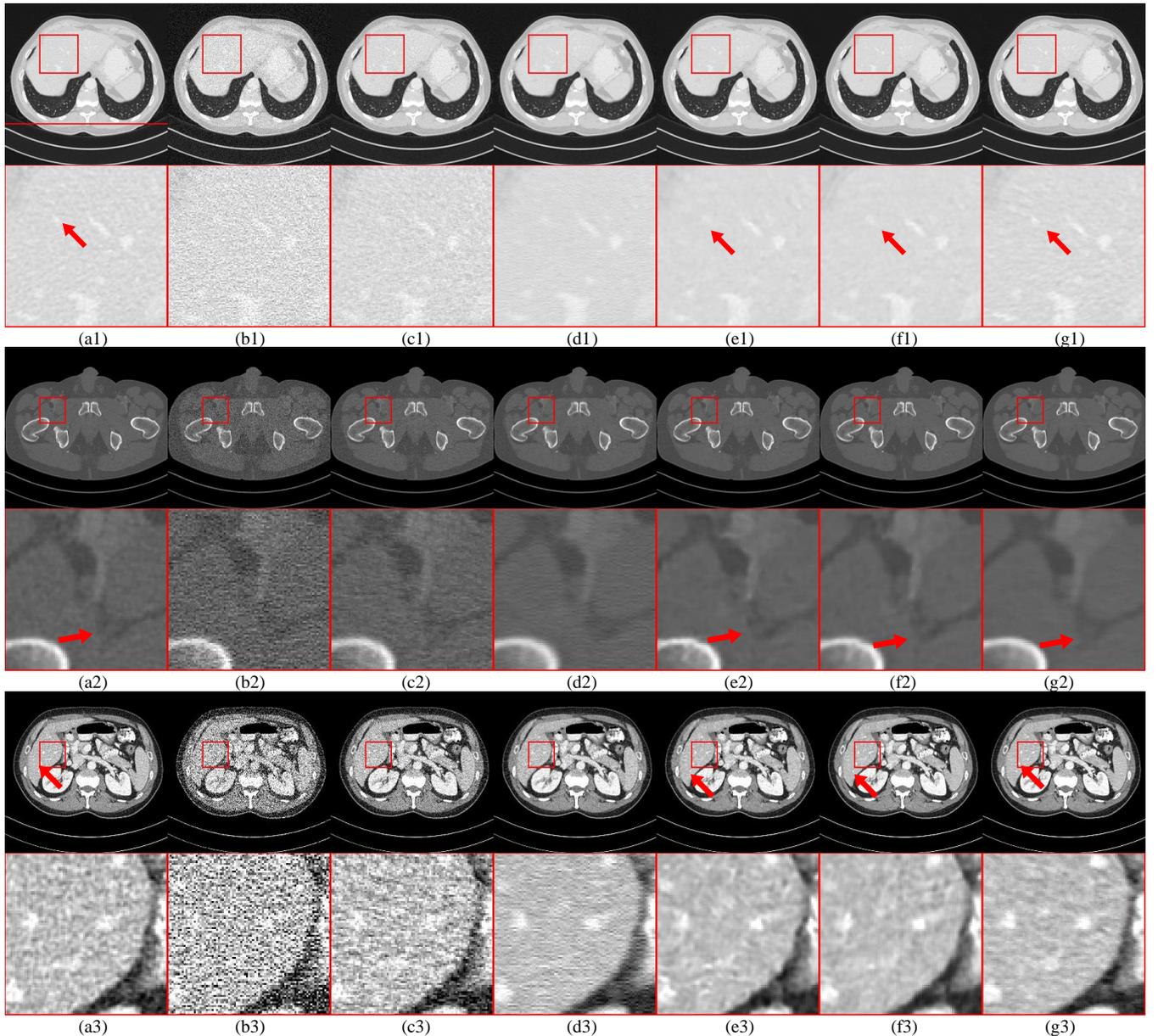

(a1) (b1) (c1) (d1) (e1) (f1) (g1)

(a2) (b2) (c2) (d2) (e2) (f2) (g2)

(a3) (b3) (c3) (d3) (e3) (f3) (g3)

**Fig. 8.** Reconstruction results of AAPM challenge data for different methods. (a1)-(a3) reference image (b1)-(b3) FBP (c1)-(c3) TV (d1)-(d3) K-SVD (e1)-(e3) RED-CNN (f1)-(f3) DP-ResNet (g1)-(g3) EASEL. The display windows are [-1150, 350] HU, [-700, 1300] HU and [-160, 240] HU, respectively.

To better illustrate the effectiveness of noise removal by EASEL, Fig. 10 plots the 1D line intensity profile passing through the red dashed line in Fig. 8(a1). It compares the same line intensity profiles from the CT image reconstructed by various methods. Through visual inspection, it is evident that the line intensity profile from our proposed method resembles most closely to the one from the normal-dose CT image. The comparison demonstrated the advantage of the proposed reconstruction method over the other reconstruction algorithms on edge preservation.

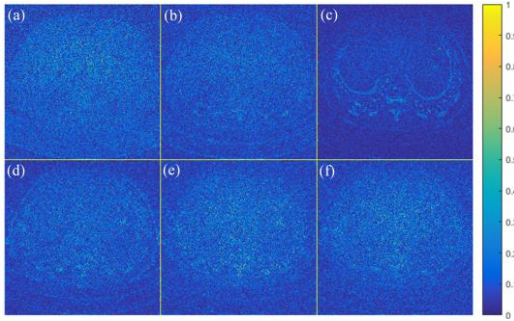

**Fig. 9.** The absolute difference images between the reference CT image and the CT images reconstructed from the different algorithms:(a) FBP (b) TV (c) K-SVD (d) RED-CNN (e) DP-ResNet (f) EASEL.

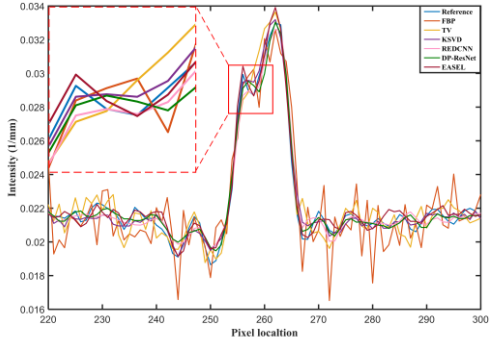

**Fig. 10.** 1D intensity profile passing through the solid red line in Fig. 8(a1). All the results in (a1)-(g1) are compared.

*CIRS Phantom Data:* For the CIRS phantom data, quantitative results reconstructed from different reconstruction methods are tabulated in Table III. It can be observed that EASEL performs better than the other methods in a trend similar to what we have seen from the reconstruction images and produces the highest PSNR. The PSNR measure for the EASEL reconstruction image improves by about 0.37dB compared to the result from DP-ResNet. In fact, compared with the end-to-end DL algorithms, the unsupervised method greatly reduces the image reconstruction effect after changing the test data.

TABLE III
QUANTITATIVE RESULTS (MEAN±STD) ASSOCIATED WITH DIFFERENT METHODS FOR THE IMAGES IN THE TESTING LD PROJECTION DATASET.

| Method | MAE | PSNR | SSIM |
| --- | --- | --- | --- |
| FBP(Ramp-filter) | 9.48±0.64 | 40.67±0.59 | 0.9436±0.0087 |
| TV | 7.64±0.41 | 42.12±0.35 | 0.9658±0.0051 |
| K-SVD | 7.01±0.19 | 42.86±0.34 | 0.9701±0.0036 |
| RED-CNN | 6.74±0.07 | 41.76±0.12 | 0.9747±0.0016 |
| DP-ResNet | **5.96±0.13** | 42.89±0.12 | **0.9811±0.0015** |
| EASEL | 6.03±0.18 | **43.26±0.08** | 0.9810±0.0022 |

To visualize the benefits of the proposed method, reconstruction images with ROIs using different methods to the ground truth (full-dose CT images) are provided in Fig. 11. Specifically, a supervised method with the network structure would cause an obvious artifact around the center of the reconstruction, whereas the similar artifact would not appear in the reconstruction of K-SVD and TV. One possible reason is that end-to-end learning models that are particularly trained for a certain task with the same data. By comparing the result of K-SVD, we find that the boundary of reconstruction image is still visible, while it is blurrier in the CT images from other methods. Moreover, DP-ResNet works a bit better than RED-CNN method but some edges and small structures are oversmoothed.

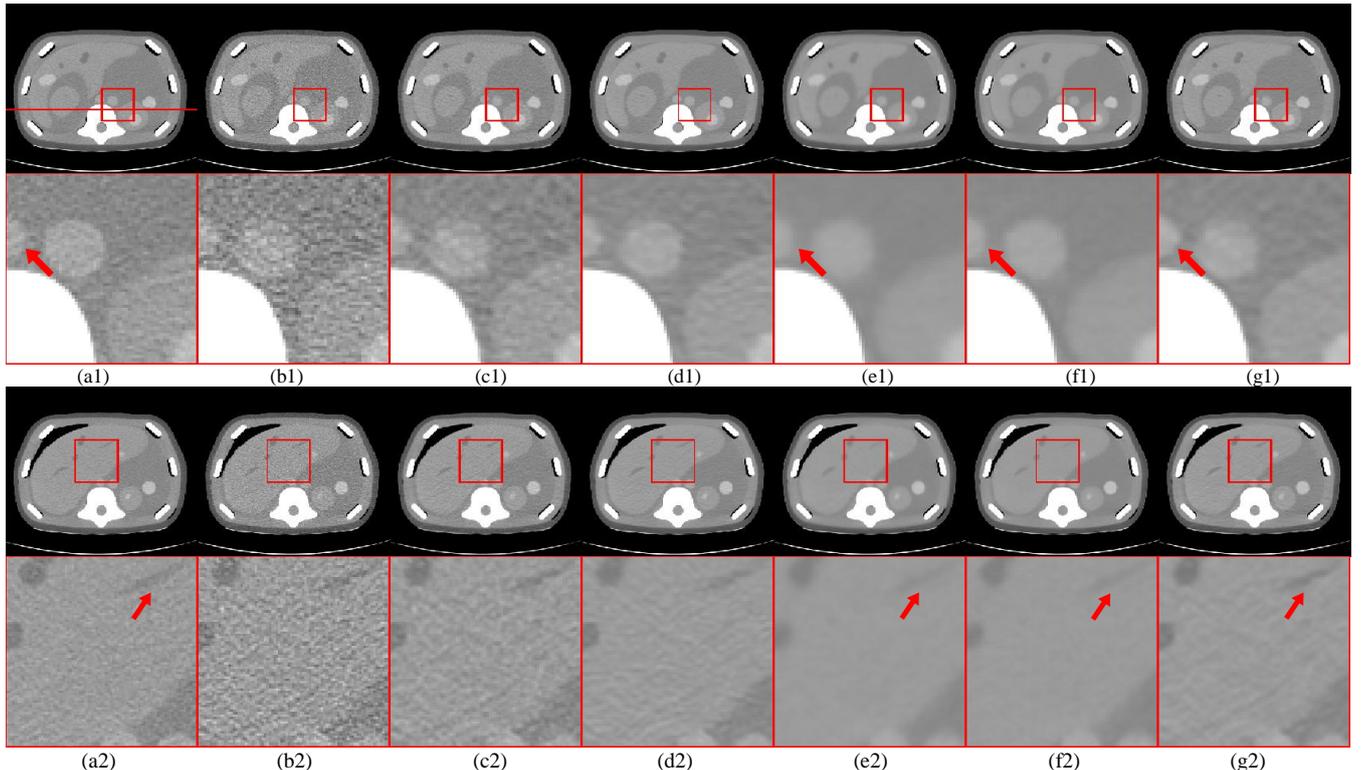

**Fig. 11.** Reconstruction results of an abdominal CT scan from CIRS phantom data using different methods. (a) reference image, (b) FBP, (c), TV, (d) K-SVD, (e) RED-CNN, (f) DP-ResNet, (g) EASEL. The display window of the full images is [-160, 240] HU.

For attaining further perspectives of our adapted EASEL for LDCT reconstruction, the residual images between the reconstructions and the reference are presented in Fig. 12, which demonstrates that the EASEL can achieve better reconstruction accuracy than that of the other algorithms on edge preservation.

To further investigate the algorithms' ability of reconstruction, Fig. 13 plots the image profiles for the six methods together with that of the ground-truth image. Intuitively, the bias can be observed more clearly in the profile plots: The pixel intensities for the DP-ResNet reconstruction better follow those of the "true" clinical image, while those for the K-SVD reconstruction are much worse than the "true" values. Moreover, the gap between the profiles of the TV method and the ground-truth shows the gigantic bias. This means that TV produces a strong bias in the reconstruction. Instead, it is obvious that the profiles for EASEL are closest to the ground-truth among the compared methods.

To sum up, in almost existing supervised DL-based methods, the network is learned by training a large amount of data that acquired with specific imaging geometry. If the observed sinograms are acquired with different low-dose scanning conditions and inconsistent with the training data, the reconstruction performance might dramatically decrease. By contrast, the proposed EASEL method largely alleviates the deficiency. In addition, we additionally calculate the average time of each iteration to facilitate the comparison of such algorithms. As can be seen from Table IV, although EASEL as an iterative algorithm requires a certain number of iterations to reach the optimal solution, its single iteration time is quite efficient in practical applications.

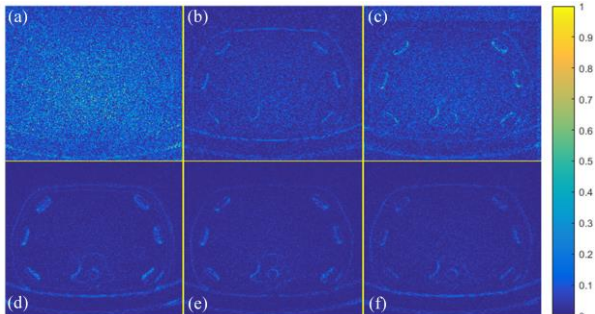

**Fig. 12.** The absolute difference images between the original CT image and the CT images reconstructed from the different algorithms:(a) FBP (b) TV (c) K-SVD (d) RED-CNN (e) DP-ResNet (f) EASEL.

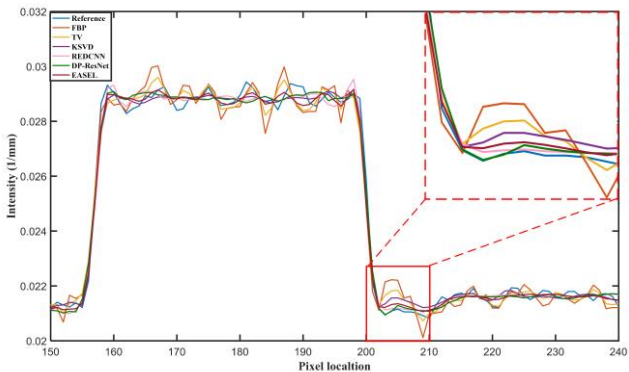

**Fig. 13.** 1D intensity profile passing through the red solid line in Fig. 11 (a1).

TABLE IV
COMPUTATIONAL COST EVALUATIONS OF VARIOUS RECONSTRUCTION METHODS. THE ORIGINAL IMAGE IS A 512×512 CT IMAGE.

| Method | Cost time (s) |
|---|---|
| FBP | 0.06 s (CPU) |
| TV | 0.26 s/Iter (CPU) |
| K-SVD | 4.77 s/Iter (CPU) |
| RED-CNN | 0.08 s (GPU) |
| DP-ResNet | 0.19 s (GPU) |
| EASEL | 0.22 s/Iter (GPU) |

### E. Variants of Hyperparameters

The hyperparameter selection is one of the most crucial factors for the image quality attainable with the proposed method. We vary one parameter at a time while keeping the others fixed at their nominal values. Additionally, the AAPM challenge data is chosen as the training and test sets. Since the $\beta$-value is the most sensitive parameter to the image quality, it is estimated by comparing the numerical indicators between the full-dose CT images and the processed LDCT images. Fig. 14 depicts the MAEs and SSIMs of the reconstructed results with different hyperparameters. It can be seen that the performance gradually improves as $\beta$-value increases. Later, the difference of performance between large $\beta$ and the peak $\beta$ vanishes. Thus, we set $\beta=150$ in our experiments.

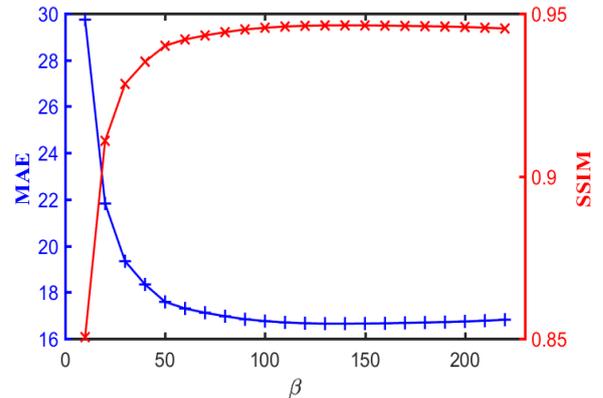

**Fig. 14.** Evolution of the regularization parameter $\beta$ for EASEL.

To examine the convergence of EASEL, the convergence tendency of SSIM curve versus iteration is plotted in Fig. 15. It can be seen that the fluctuation of the curve is not obvious as the iteration increases. Besides, it is evident that EASEL is effective for noise and artifact suppression of LDCT images after 1500 iterations. Therefore, EASEL has a reasonably fast convergence rate.

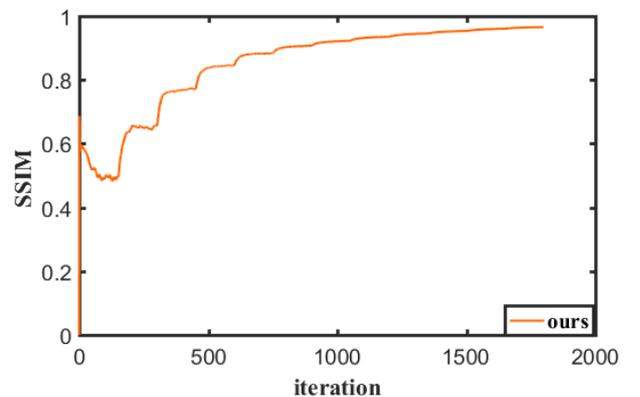

**Fig. 15.** Convergence tendency of EASEL in LDCT reconstruction.

In addition, the reconstruction results with regard to the number of input channels of the network $s_\theta(x)$ are inves-tigated in Table V. It is predictable that, as the channel number increases, the representation ability of the prior leverages. Subsequently, the performance will be improved. Considering the computational complexity, the channel number is set to be 10.

TABLE V
THE IMPACT OF CHANNEL NUMBER ON LDCT RECONSTRUCTION.

| Channel | 3 | 5 | 10 | Naïve RefineNet |
|---|---|---|---|---|
| MAE | 15.70 | 15.81 | **15.58** | 15.84 |
| PSNR | 41.26 | 41.37 | **41.50** | 41.27 |
| SSIM | 0.9525 | 0.9533 | **0.9539** | 0.9520 |

To verify the superiority of EASEL as a tool of the unsupervised learning methodology, we use ESAEL and DP-ResNet to reconstruct LDCT images with different noise levels, and do not re-train the model. The experimental numerical indicators are shown in Table VI. As seen, EASEL is a bit superior to DP-ResNet for reconstructing the noisy data under noise level $b_i=1e5$. However, if we use the same trained DP-ResNet model to tackle the noisy data under noise level $b_i=1e4$, the performance of DP-ResNet degrades heavily. While our method alleviates the remedy much and the performance still retains good behavior, the PSNR of EASEL is 3.7db higher. The numerical difference can be clearly shown in the reconstruction image in Fig. 16. As observed, the reconstruction in Fig. 16(d) is more helpful for clinical diagnosis than that in Fig. 16(c). Overall, the proposed method EASEL can handle a broad range of noise strengths.

TABLE VI
QUANTITATIVE RESULTS (PSNR/SSIM) ASSOCIATED WITH DIFFERENT NOISE LEVEL IN AAPM TEST DATASET.

| Method | $b_i=1e5$ | $b_i=1e4$ |
|---|---|---|
| DP-ResNet | 42.45/0.9609 | 34.97/0.8547 |
| EASEL | **42.57/0.9621** | **38.67/0.9228** |

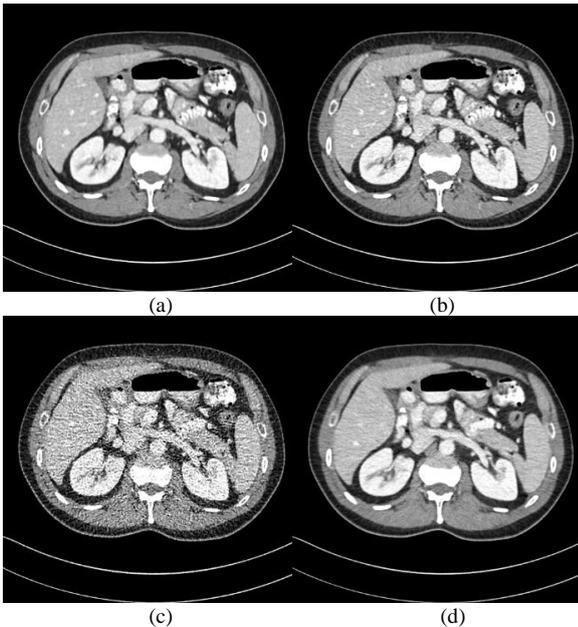

**Fig. 16.** Reconstructed results for different noise levels. (a) DP-ResNet result with $b_i=1e5$, (b) EASEL result with $b_i=1e5$, (c) DP-ResNet result with $b_i=1e4$, (d) EASEL result with $b_i=1e4$.

In addition, we compare the proposed method with the GAN-based method [75]. Table VII gives the evaluation index of the reconstruction result. From the table, one can see that EASEL is superior to the GAN-based algorithm. Besides, some visualizations of the reconstruction results are shown in Fig. 17. It can be observed that the reconstruction results of the GAN-based method still retain some noise.

TABLE VII
QUANTITATIVE RESULTS (MEAN±STD) IN THE AAPM DATASET.

| Method | MAE | PSNR | SSIM |
|---|---|---|---|
| GAN-based | 16.43±1.00 | 41.02±0.54 | 0.9468±0.0068 |
| EASEL | **15.44±1.17** | **41.58±0.67** | **0.9539±0.0067** |

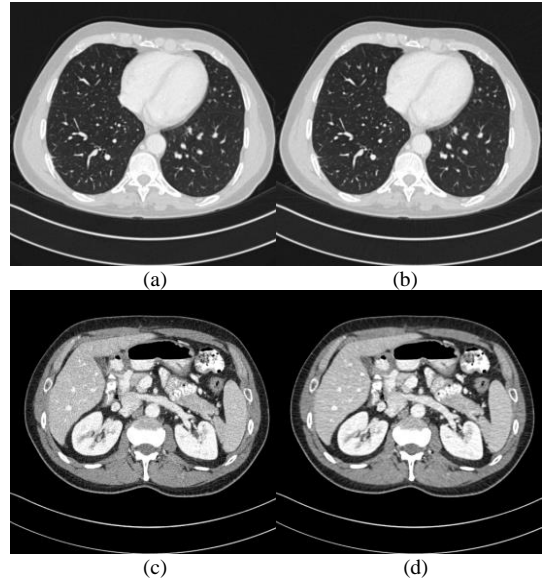

**Fig. 17.** Reconstructed results from different methods. (a)(c) GAN-based, (b)(d) EASEL.

To compare the visual quality of different algorithms, we have added a couple of medical images with uncommon pathologies in visual comparison experiment. Fig. 18 shows that the EASEL reconstructed low-dose CT images suffer less from noise and artifacts with a better tissue identification. The results of the EASEL preserved the textures of the hemangioma tumor such that a better determination of the location of the lesion was possible, as highlighted by the zoom-in area.

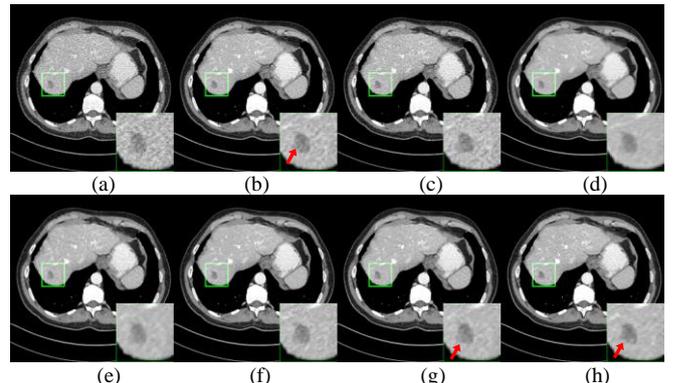

**Fig. 18.** Reconstruction results by different methods. (a) FBP, (b) Reference image, (c) TV, (d) K-SVD, (e) RED-CNN, (f) GAN-based, (g) DP-ResNet, (h) EASEL.

### F. Analysis of Loss Function

The loss function, despite being the effective driver of the network's learning, has attracted little attention within the image processing community [65]. The choice of the cost function generally defaults to be the squared $L_2$ norm of the error. In this study, we bring attention to loss function for

LDCT reconstruction. Specifically, the mean squared error $L_2$ penalizes larger errors, but it is more tolerant to small errors, regardless of the underlying structure in the image. By contrast, loss function with $L_1$-norm does not over-penalize larger errors. Consequently, they may have different convergence properties. Inspired by this observation, we test whether a different local metric such as $L_1$ can produce better results with the state-of-the-art metrics for image quality. The performance of several losses (i.e., $L_1$, $L_2$, $L_1+L_2$) is recorded in Table VIII. As can be observed, the reconstruction quality varies scarcely with the loss functions.

$L_2$-norm is arguably the dominant error measure across very diverse fields, from regression problems, to pattern recognition, to signal and image processing. The main reason for its popularity is the fact that it is convex and differentiable—very convenient properties for optimization problems. Meanwhile, it provides the maximum likelihood estimate in case of independent and identically distributed Gaussian noise, to the fact that it is additive for independent noise sources. Overall, considering the case of our work that performs noise distribution mapping, i.e., the distribution obeys normal distribution and has the same mathematical expression as $L_2$-norm, we choose $L_2$-norm as loss function to avoid the selection of complex parameters.

TABLE VIII
THE IMPACT OF LOSS FUNCTIONS ON LDCT RECONSTRUCTION.

| Loss function | $L_1$ | $L_2$ | $L_1+L_2$ |
|---|---|---|---|
| MAE | 15.78 | 15.84 | **15.70** |
| PSNR | 41.24 | **41.27** | 41.19 |
| SSIM | 0.9482 | **0.9520** | 0.9515 |

## V. CONCLUSIONS AND DISCUSSIONS

In this work, a novel iterative reconstruction framework, coined EASEL, was proposed. By annealing the gradient of data of density, elaborate prior knowledge from generative models was incorporated into the iterative procedure. Experimental results on two public datasets demonstrated the feasibility and efficiency of EASEL for LDCT imaging problem, improving image quality and avoiding noise effect. On the one hand, to effectively extract image features at multiple scales, we self-copied the LDCT image into ten channels, and then the generative network used the ten components as input. Moreover, the modification of network architecture was used in our experiments to better combine high-dimensional information for LDCT reconstruction. On the other hand, a novel generative model where samples were produced via Langevin dynamics using gradients of the data distribution estimated with DSM was utilized for LDCT reconstruction. In addition, the distance between the reconstructed images and the learned manifold was minimized along with the data fidelity by SQS algorithm during iterative reconstruction. As a result, the reconstructed images become closer to the reference data.

In future study, we will extend our model to find the image similarity search on latent space over huge clinical image dataset and deal with more challenging tasks. The most significant requirement will be the availability of a larger multi-GPU server architecture. Experiments that use such a computational infrastructure are under-way. Besides, we would also like to integrate the deep gradient priors of generative model into an iterative reconstruction pipeline to mitigate possible image feature suppression imposed by the network.

ACKNOWLEDGMENTS

The authors sincerely thank the anonymous referees for their valuable comments on this work. They would also like to thank Dr. C. McCollough of the Mayo Clinic, Rochester, MN, USA, for providing clinical projection data, as agreed under the American Association of Physicists in Medicine.